\documentclass[11pt, a4paper, pdftex]{article}
\usepackage[utf8]{inputenc}		%\usepackage[applemac]{inputenc}
\usepackage[T1]{fontenc}

\usepackage{amsmath}
\usepackage{amssymb}
\usepackage{setspace}
\usepackage{latexsym}
\usepackage{fancyhdr}
\usepackage{lastpage}
\usepackage{times}

\usepackage{notations} 
\usepackage{myAlgorithm}

\usepackage[pdftex]{graphicx}

\usepackage[english]{babel}

\usepackage{authblk}

\graphicspath{
	{my_figures/}
}

\newcommand{\pa}{\paragraph{}}

\def\og{\leavevmode\raise.3ex\hbox{$\scriptscriptstyle\langle\!\langle$~}}
\def\fg{\leavevmode\raise.3ex\hbox{$\scriptscriptstyle\rangle\!\rangle$}}

\begin{document}

\title{An alternative approach for the convolution in time-domain: the taches-algorithms}
\author[1, 2]{Laurent Millot \thanks{Electronic address: l.millot@ens-louis-lumiere.fr; corresponding author}}
\author[1, 2]{Gérard Pelé}
\affil[1]{Institut d'esthétique, des arts et technologies - UMR 8153 (CNRS/Université Paris 1/MENRT), 
Université Paris 1, 47 rue des bergers, Paris, 75015, France}
\affil[2]{ENS Louis-Lumière, Noisy-le-Grand, 7 allée du promontoire, 93160, France}

\date{11/03/2008}

\maketitle

%\doublespacing

%%%%%%%%%%%%%%%%%%%%%%%%%
	\begin{abstract}
We present an alternative temporal approach for convolution, providing a new algorithm, called the taches-algorithm. Based on interferences between the successive delayed and amplified output signals associated respectively with the impulses constituting the input signal, the taches-algorithm can give access immediately to the new output sample and have a low latency response even without using vector-based optimisation of the calculation. With the taches-algorithm it seems easy to change (even in real-time) the impulse response while running the calculation, simply by updating the impulse response to use it for next samples, a task rather difficult to achieve using FFT convolution. Real-time audio demonstrations using notably Pure Data and simple explanations of the taches-algorithm will be given.

Paper 7412 presented at the 125th Convention of the Audio Engineering Society, Amsterdam, 2008

	\end{abstract}

\section{Introduction}
\pa In this paper we consider the validity of an \textit{a priori} not used temporal algorithm to perform the calculation of convolution. We chose  to name this algorithm the taches-algorithm due to its physical explanation. This taches-algorithm is a FFT-free algorithm as it is exposed in the following.

\pa First, we explain why we have chosen to propose a paper on this subject.

\pa Second, starting from a physical experiment chosen as a reference by Millot, we show how one can refind the time-domain convolution well-known formula but with a rather different point of view: not a sample-by-sample point of view but a whole signal approach.

\pa Third, we give two examples of pseudo codes for the taches-algorithm, one using the vectors operations, and the other using simple buffers. Some clues for optimization of the taches-algorithm are also reviewed in this section.

\pa Then, using a C++ code inside a Xcode projet, we study the influence of the filter impulse length in order to determine for several different computers the limit length which would permit real-time calculation, the influence of the input buffer size over the calculation time and the question of the quantization errors. 

\pa Finally, we propose some various potential uses for such an algorithm.

\section{Why exposing such an algorithm}
\pa In the 122th AES Convention in Vienna, during the session "Spatial Audio Perception and Processing, part 2", Millot attended to the presentation of paper  7021 \cite{Stewart:07}, given by Stewart and dealing with hybrid artificial reverberation. The main idea of the presented algorithm is using temporal convolution for the early reflections and a filter bank to process the late reflections. One of the main difficulties with this algorithm is related to the spectral analysis of the impulse response in order to determine the para\-meters defining the filter bank. Indeed it seems rather difficult to find a simple and automatic criterion to define how to realize the early/late reflections separation. 

\pa During the discussion, it appears that the temporal algorithm which is presented in this paper was unknown from the audience while it may either be able to give some clues to solve the problem of the filter bank parameters definition or constitue an alternative way of research. Discussions after the end of this session with other participants convinced Millot that it may be useful to present this temporal algorithm and some ways to code it.

\pa This algorithm, the so-called taches-algorithm, is the natural way used by Millot \cite{Millot:08} to introduce convolution for sound engineering students at the ENS Louis-Lumière and the Formation Supérieure aux Métiers du Son (CNSMDP) for almost 10 years. A similar introduction, without the physical reference discussed in next section, can be found in \cite{Orfanidis:96, Smith:99, Smith:07} for instance but this way to think the convolution does not seem to be used further by these authors to program and process the convolution in time-domain. 

\pa We have been testing such a FFT-free algorithm within a running project to propose an alternative characterization method for audio installations using the IDS analysis \cite{Millot:04, Millot:07} and, above all, music or speech as measurement stimuli. And, at this point, the taches-algorithm appears as a rather smart and efficient tool to perform the convolution part of the IDS analysis. 

\pa Actual computers may be able to permit convolution using the taches-algorithm with small to acceptable calculation times, as we work for the moment off-line. Calculation times were "unacceptable" either in 2003 when Samuel Tracol, student at the FSMS, proposed a C coding of Millot's proposal or when Benjamin Bernard \cite{Bernard:03} tested the taches-algorithm within the hardware Pyramix digital workstation. But, the increase of the power computing makes possible the use of the taches-algorithm today and, moreover, tomorrow. 
  
\pa Another reason is the fact that time-domain convolution is a subject quite often fastly presented in Signal Processing courses as one may think that the normal way to perform fast convolution is using fast Fourier transforms (FFT). So, it is common to find the convolution introduced directly using its mathe\-matical "ugly" derivation, even in books for beginners \cite{Zolzer:02}: 
\begin{equation}
s[n]=\sum_{m=-\infty}^{+\infty}e[m].h[n-m]
\label{eq:ConvoBase}
\end{equation}
which, according to us, does not permit to really understand what is a convolution. 

\pa Next step is often to look at the simple way to perform convolution in frequency-domain using the transformation of convolution in a simple multiplication of the input and filter impulse response transforms:
\begin{equation}
S[k]=H[k].E[k]
\label{eq:ConvoFFT}
\end{equation}
where, here, $S[k]$, $H[k]$ and $E[k]$ are respectively the discrete Fourier transforms (DFT) of the output $s[n]$, filter impulse response $h[n]$ and input $e[n]$ signals.

\pa In Eq. (\ref{eq:ConvoFFT}), we use the following definition for the DFT $E[k]$, for instance, of the input signal $e[n]$ of length $N_e$:
\begin{equation}
E[k]=\sum_{n=0}^{N_e-1}e[n].\exp(-2\pi\mathrm{j}\frac{kn}{N}).
\label{eq:DefDFT}
\end{equation}

\pa And, in order to increase the speed of the convolution, the normal process is to use an optimized version of the calculation of the DFT, the so-called Fast Fourier Transform (FFT), and of the inverse FFT (IFFT). In such a case, the fast convolution is then realized following the processing scheme described in Fig. 1, and with optional zero-padding of input and impulse response signals to consider power of 2 signal lengths before calculating the FFTs.

\begin{figure}[h!t!]
\begin{center}
\includegraphics[width=12cm]{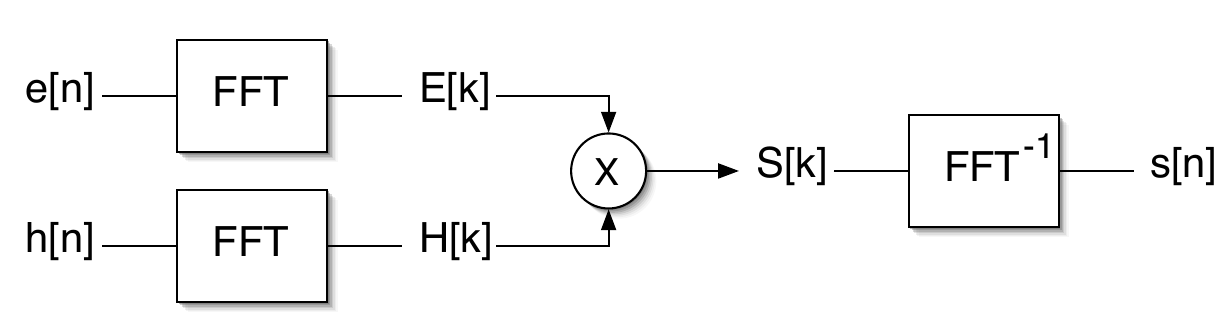}
\end{center}
\nobreak
\caption[1]{Fast convolution algorithm using FFT and IFFT (FFT$^{-1}$ block-diagram) in order to substitute to a rather complex temporal calculation of the convolution using Eq. (\ref{eq:ConvoBase}) a term-by-term multiplication of both FFTs $E[k]$ and $H[k]$.}
\end{figure}

\pa So, as the time-domain convolution is often consi\-dered as useless or too much time consuming, the effort is focused on the study of frequency-domain approach and as a consequence of the fast convolution using FFTs.

\pa But, we will try to underline in this paper why we think that the taches-algorithm could be a convin\-cing alternative for either off-line or even real-time processing for a growing set of situations. 

\section{Physical origin and principle}
\subsection{Physical origin}
\pa To explain the convolution, Millot \cite{Millot:08}, used a classical experiment in Optics that most of the students have encountered in their former educational cursus. 

\pa This experiment, whose experimental set-up is given in Fig.~2, consists in studying how a converging thin lens transforms a punctual source, located far away from the lens in the object space, within the image space. It is then easy to underline that the observation screen must be placed at the image focus in order to get the best image of the image of the source given by the converging thin lens.

\begin{figure}[h!t!]
\begin{center}
\includegraphics[width=8cm]{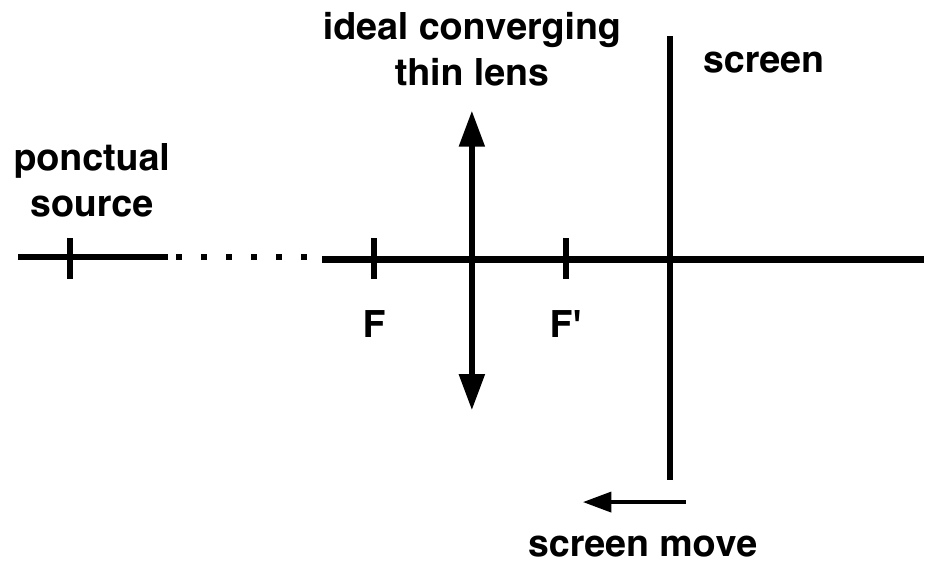}
\end{center}
\nobreak
\caption[2]{Classical experiment in Optics: a punctual source located \textit{at infinitum} in the source space, ideal converging thin lens, a screen one moves within the image space in order to find the best location to observe the image of the punctual source.}
\end{figure}

\pa From a theoretical point of view, if we assumed that the source is really punctual, located \textit{at infinitum}  in the source space and that the thin lens is also ideal, the image given by the thin lens of the punctual source is the Airy spot, a "tache" in French. 

\pa The theoretical Airy spot and its profile over a dia\-meter of the lens are given in Fig.~3. This profiles coincides with a sinc function as, in theory, the ideal thin lens is a perfect low-pass filter for light.

\begin{figure}[h!t]
\begin{center}
\includegraphics[width=6.5cm]{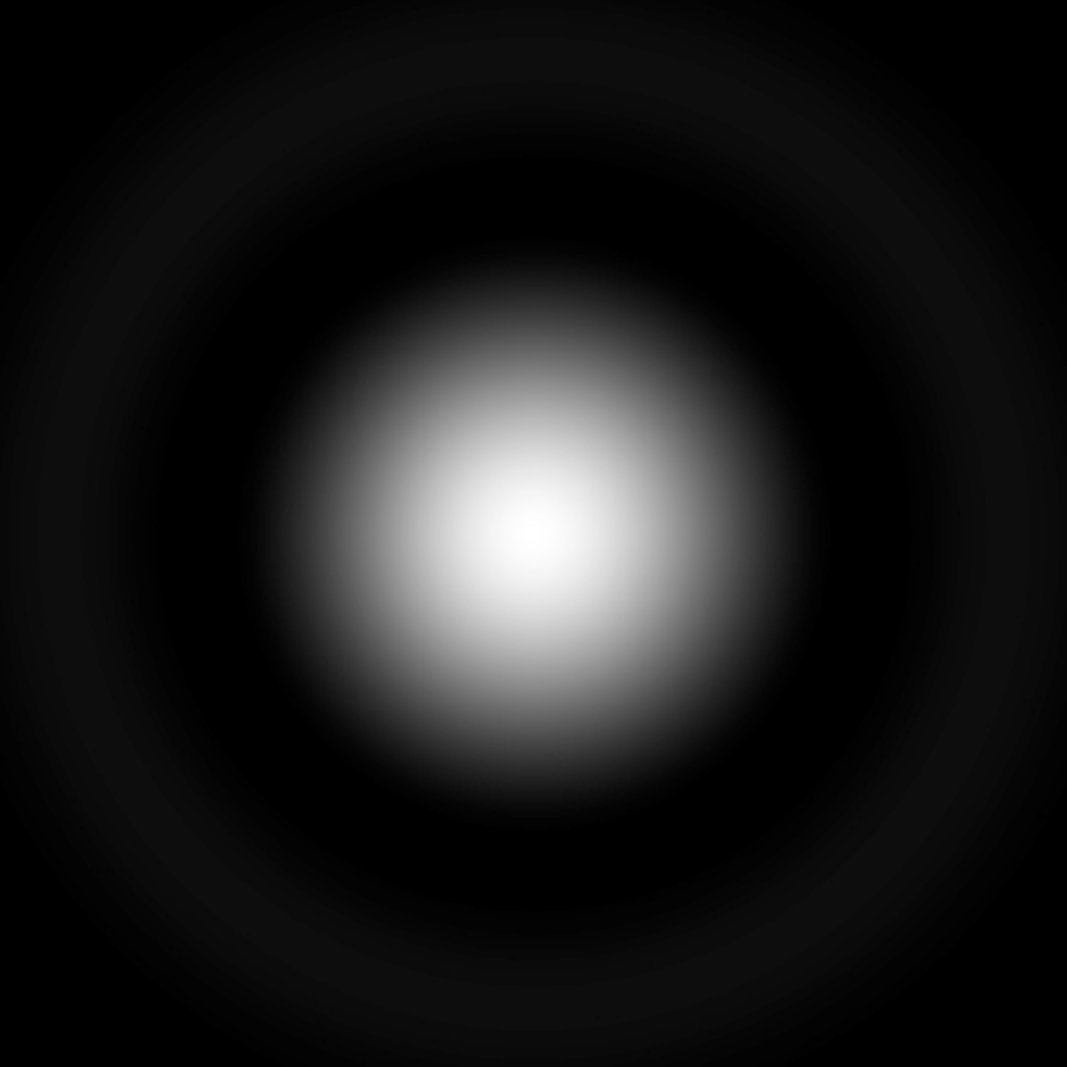}
\includegraphics[width=7.5cm]{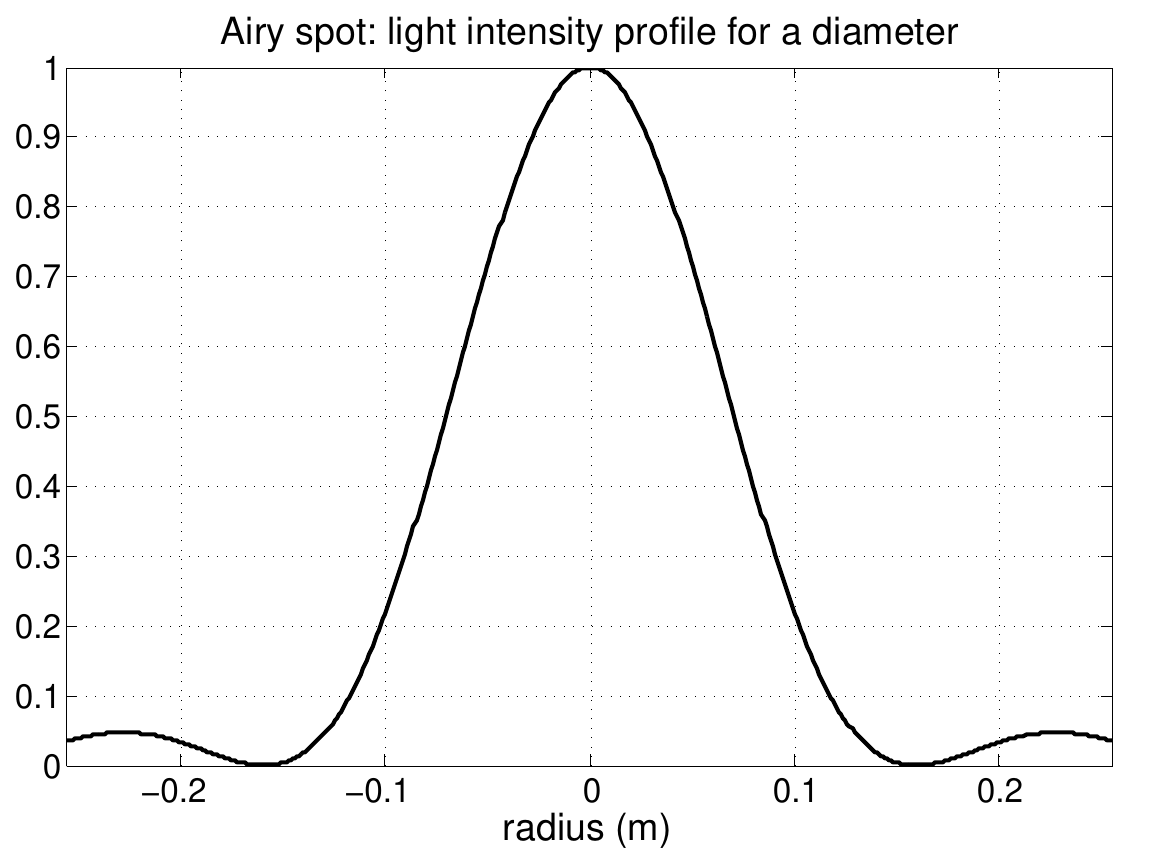}
\end{center}
\nobreak
\caption[3]{(left) Airy spot and its theoretical profile versus a diameter of the thin lens (right) which coincides with the square of a sinc function.}
\end{figure}

\pa A spatial displacement of the punctual source out of the thin lens axis gives an opposite displacement of the tache on the observation screen and a modulation of the source power would also give a similar  modulation of the tache on the screen.  And, considering a rather complex source with a finite spatial extension, the related image on the observation screen is the sum of all the taches associated with each of the punctual sources composing the image \textit{at infinitum}. Such a physical phenomenon is classically explained assuming the linear superposition or interferences of waves.

%\clearpage
\subsection{Principles of the taches-algorithm}
\pa We can use this physical phenomenon to refind the principle of convolution. 

\pa Indeed, the Airy spot corresponds to the response of a filter, the ideal thin lens, to an impulse related to the punctual source located \textit{at infinitum}. So, the Airy spot or tache coincides with the principle of impulse response of a filter in Signal Processing. 

\pa Following this analogy, we can suggest that a digital filter (we focus on digital filters but the method is the same for analog ones) gives a tache, the impulse response $h[n]$, when excited by an ideal unitary impulse $\delta[n]$: the filter cannot gives an exact image of the excitation impulse $\delta[n]$ and, then, provides as output an approximation or tache which is called the impulse response $h[n]$.

\pa Considering that the filter has a linear and time-invariant behavior (LTI filter), the response to a scaled and delayed impulse excitation at the input is then a tache or impulse response delayed and scaled exactly by the same values. 

\pa So, if we consider as input excitation the signal $e[n_0].\delta[n-n_0]$, the related filter output is the tache $e[n_0].h[n-n_0]$: a version of the impulse response delayed by $n_0$ samples and scaled by the $e[n_0]$ factor.  And, as we can decompose any input signal $e[n]$ as a succession of delayed and scaled impulses $e[n_0].\delta[n-n_0]$, the output signal $s[n]$ coincides with the superposition, interference or sum of all the partial outputs, or taches, $e[n_0].h[n-n_0]$ where $n_0$ varies over the range of the samples composing the input signal. 

\pa From a mathematical point of view, we are then able to derive the formulation giving the output \textbf{signal} or running sample, according to the point of view one may adopt, as a sum of each tache signal corresponding to the delayed and scales versions of the characteristic filter tache or impulse response. Indeed, as we can write the input signal as :
$$e[n]=\sum_{m=N_1}^{N_2}e[m].\delta[n-m]$$
where $N_1$ and $N_2$ are respectively related to the first and last samples of the input, we can simply synthesize the output signal with:
\begin{equation}
s[n]=\sum_{m=N_1}^{N_2}e[m].h[n-m].
\label{eq:Convo}
\end{equation} 

\pa According to the respective values of $N_1$ and $N_2$ we refind the classical mathematical formulation of time-domain convolution, given by Eq. (\ref{eq:ConvoBase}).

\pa But, it seems rather important to underline the fact that we consider Eq. (\ref{eq:Convo}) as a signal definition for the output, not as a mathematical formula to calculate the samples of the output signal sample-by-sample.

\pa The difference of point of view is essential as we will see, in the following, using the example of the input and filter impulse response defined by Fig.~4.

\begin{figure}[h!t!]
\begin{center}
\includegraphics[width=7.5cm]{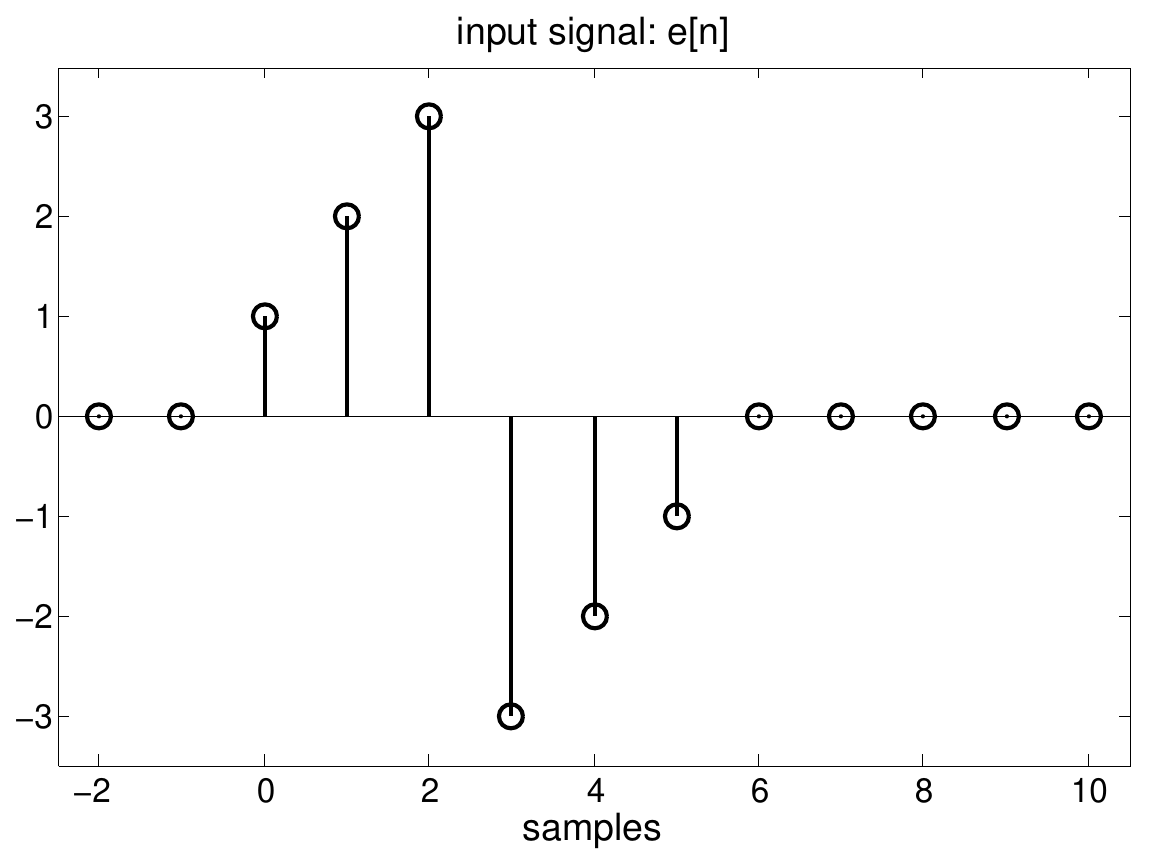}
\includegraphics[width=7.5cm]{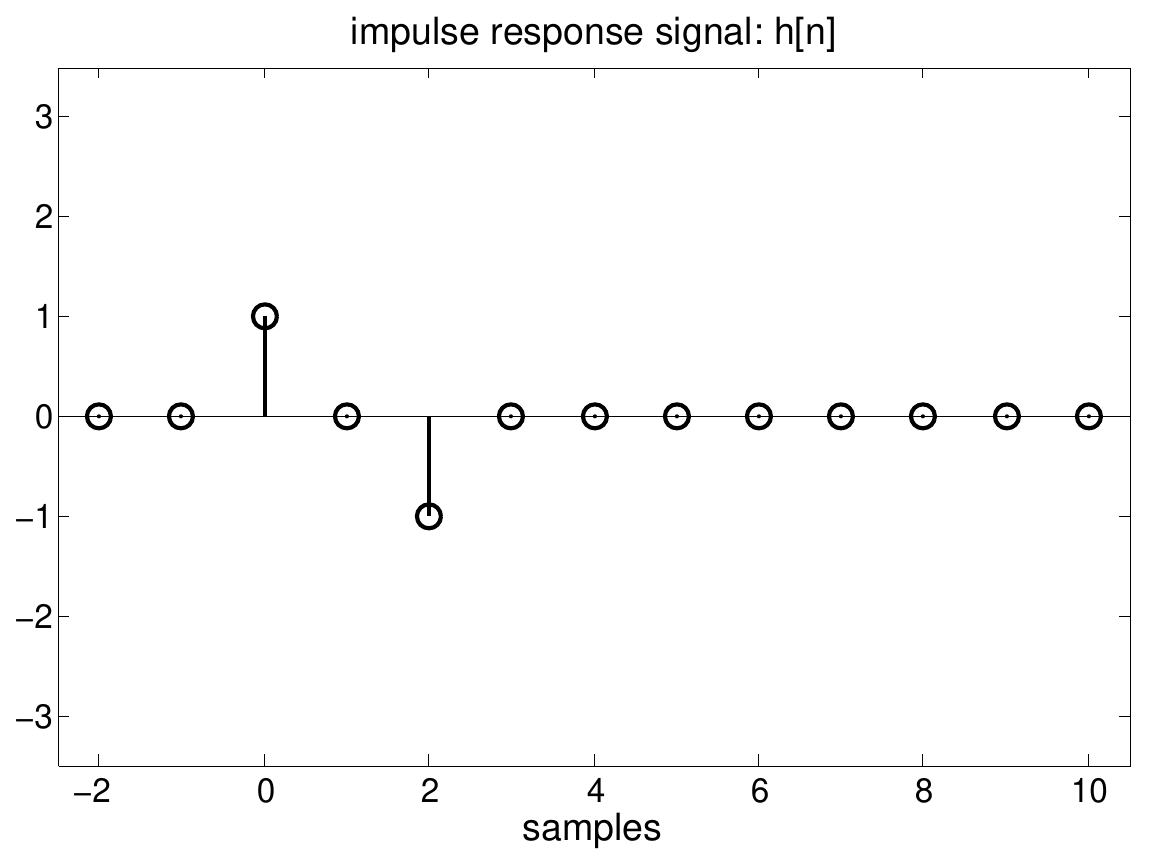}
\end{center}
\nobreak
\caption[4]{(left) Input signal and (right) filter impulse response used to explain the tache-algorithm principle.}
\end{figure}

\pa Indeed, we can synthesize the output signal consi\-dering the sum of 6 successive, delayed one sample by one sample and scaled by the values of the input signal, taches or versions of the impulse response. The whole process is described by Fig.~5 where we have arranged the successive taches vertically in order to figure such a classical addition, but for an 6-signal addition. 

\begin{figure}[h!t!]
\begin{center}
\includegraphics[width=12cm]{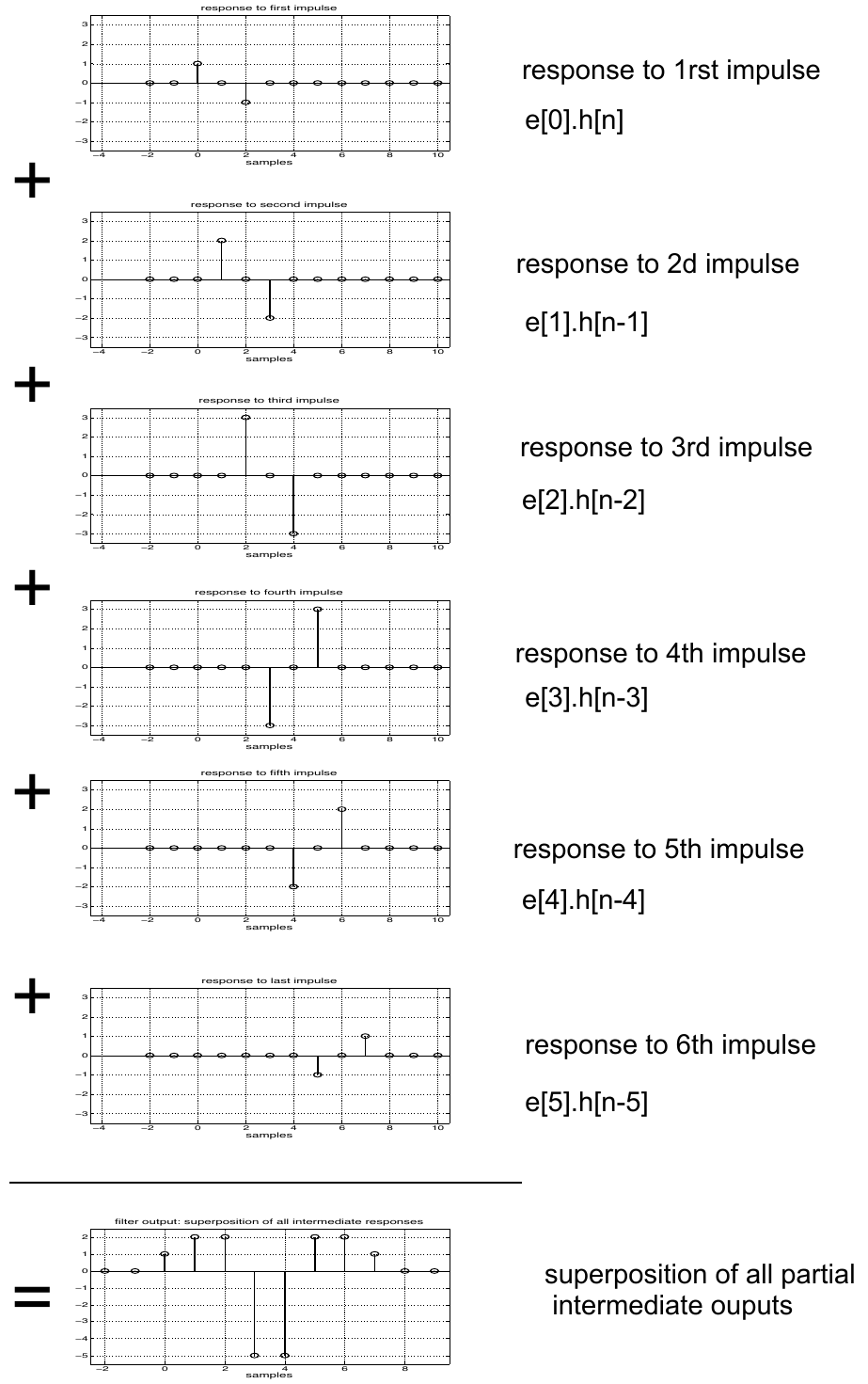}
\end{center}
\nobreak
\caption[5]{Taches-algorithm principle: each input value generates a scaled and delayed version of the impulse response and the global output is the result of linear superposition of all the intermediate outputs.}
\end{figure}

\pa As we both know the input and impulse response signals we are in a peculiar situation where we can exploit a mathematical property of the time-domain convolution formulation: commutativity.

\pa We are then able to exchange the role of the input and impulse response signals, that is to say consi\-dering input signal as filter impulse response and filter impulse response as input. In the studied case, this is quite interesting because input signal is  6-sample long while impulse response is 3-sample long. It simply means that we can calculate the output signal just considering the superposition of 3 delayed and scaled versions of a new tache, the input signal, where the delays and scales factors correspond to the samples of the impulse response. This simpler process is illustrated on Fig.~6.  

\begin{figure}
\begin{center}
\includegraphics[width=12cm]{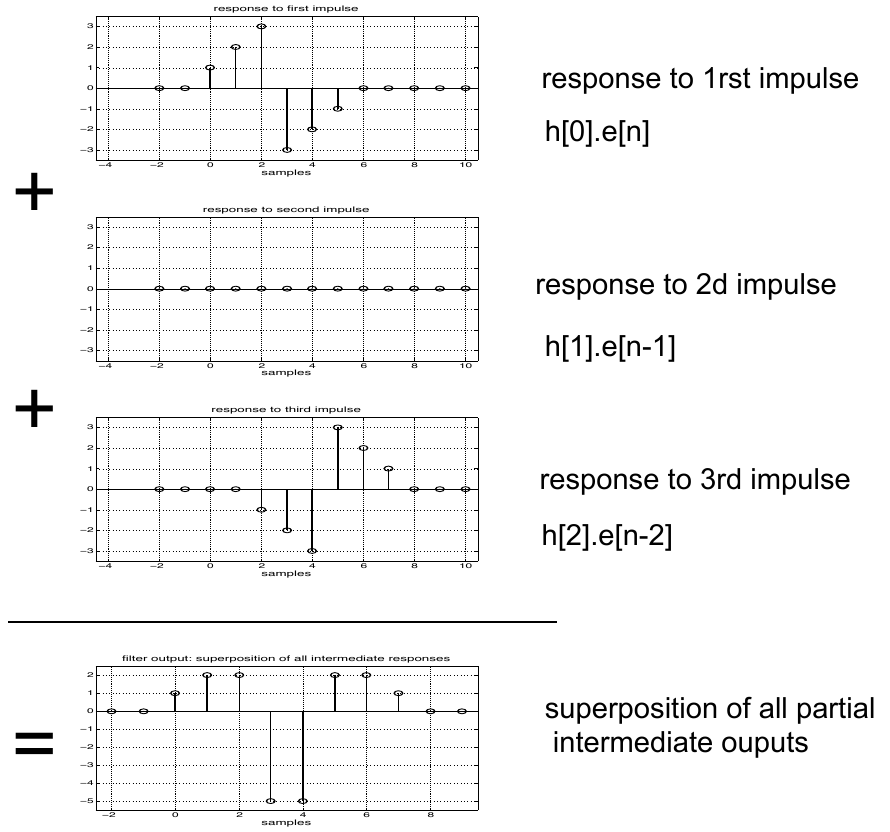}
\end{center}
\nobreak
\caption[6]{Using impulse response as input and input as impulse response can need less but longer intermediate outputs to be superimposed in order to synthesize the global output. But both impulse response and whole input must be completely available to use this solution, which means an off-line convolution.}
\end{figure}

\pa This optimization is only possible when the impulse response signal is shorter than the input one and, above all, when we completely know the input signal. This optimization can then be considered, notably, in case of off-line signal processing. For real-time applications, this optimization probably could not be used because we do not know the whole input signal.

\pa One may also note that we could think about a simple optimization each time we have a zero value for the "input" current sample. Indeed, a zero value do not generate another tache to take into account.   

\pa This principle to calculate the output signal can be refound in literature \cite{Orfanidis:96, Smith:99, Smith:07} without any clearly defined and/or explicit name ("LTI form" in \cite{Orfanidis:96} who introduces several other ways to consider the time-domain convolution) to explain the convolution, but, it seems that it has not been used to perform time-domain convolution. In fact, time-domain convolution seems always considered, when programming, from the "direct form" of convolution \cite{Orfanidis:96} that is to say as sample-by-sample routine with a summation over the samples of the impulse response.

\pa In the following, we propose to consider some ways to compute the signal-superposition approach, the so-called taches-algorithm.   

\section{Coding the taches-algorithm}
\pa The way to code the taches-algorithm is directly related to the nature of most efficient mathematical operations within a given language.

\pa The first pseudo code we propose directly relies on the vector-calculation conception, and would be adapted to be used within Matlab or Scilab for instance. Moreover, this first code is quite close to the signal approach as we can simply multiply a signal stored in an array, here the impulse response $h$, and add it to another array corresponding to the remaining "tracks" of the past taches related to the delayed and scaled impulse responses which occurred in the past.

\subsection{Vectorized version of the pseudo code}
\pa The vector-based pseudo code is then the following:
%%%%%%%%%%%%%%%%%%%%%%%%%%%%%%%%%%%
\begin{figure}[h!t]
 \begin{algorithm}{$\Algo{TachesAlgorithmVect}(e,h)$} 
	 \Aitem Ne $\setto $ size(e)  
	 \Aitem Nh $\setto $ size(h)  
	 \Aitem Nh2 $\setto $ Nh - 1  
	 \Aitem r(1..Nh) $\setto $ 0  
	 \Aitem s $ \setto$ ""  
	 \Aitem \For $n \setto 1$ \To Ne  
	 \Aitem \mttt r(..) $\setto $ r(..)+ e(n) * h(..)
	 \Aitem \mttt s $\setto $ s \^{ } r(1)
	 \Aitem \mttt r $\setto $ r(2..Nh) \^{ } 0
	 \Aitem s $\setto $ s \^{ } r(1..Nh2)
	 \Aitem \Return s
\end{algorithm}
\end{figure}
%%%%%%%%%%%%%%%%%%%%%%%%%%%%%%%%%%%

\pa The residue r corresponds to the remaining "tracks" of the past taches which is updated during each loop by adding the new tache associated with ano\-ther occurrence of the impulse response related to and scaled by the current input sample value (line 7). 

\pa In this function, the output signal is built sample-by-sample, in the "for" loop, by concatenation of the current value of the output signal (line 8) and with an initialization to an empty array (line 5). One of the advantages of the taches-algorithm is the fact that we immediately access the value of the current output, without having to wait for the processing of a  whole block as in the classical fast convolution using FFT and IFFT. This advantage will be even more obviously underlined by the buffer-based version of the pseudo code given in the following.

\pa Another curiosity is the reference to temporal overflow defined by line 10. In fact, as each input sample generates a new occurrence of the tache or impulse response, scaled by its value, the output signal is longer than the input one: for instance, the last sample or impulse of the input signal generates a last occurrence of a scaled impulse response, beginning with the last sample of the input but finishing $N_h-1$ samples later if $N_h$ is the impulse response length. Thus, after the end of the "for" loop, we have to concatenate the residue array, but without the last element which is a zero, to complete the output signal with the so-called temporal overflow (line 10).

\pa One may experiment that using basis function of fast convolution within Matlab or Scilab, the output signal calculated has exactly the same size as the input one. Indeed, the temporal overflow is not given by such functions which can be annoying when using filters likes allpass filters or filters with great group delay for peculiar frequencies. If one is interested in calculating the whole signal, with the temporal overflow, he would better program his own version of the fast convolution. And, the same problem may arise from using a black box fast convolution program or compiled application without testing it or verifying the choice made for the temporal overflow in the documentation.

\subsection{Buffer-base pseudo code}
\pa For this case, we process the convolution of the input block-by-block so we calculate partial output blocks ps from partial input blocks pe. The  related pseudo code for the taches-algorithm is then:
%%%%%%%%%%%%%%%%%%%%%%%%%%%%%%%%%%%
\begin{figure}[h!t]
 \begin{algorithm}{$\Algo{TachesAlgorithm}(pe,h, Nb)$} 
	 \Aitem r\_index $\setto $ 1  
	 \Aitem Nb $\setto $ min(Nb, size(pe))  
	 \Aitem Nh $\setto $ size(h)  
	 \Aitem \For n = 1 \To Nh
	 	\Aitem \mttt r(n) $\setto$ 0
	 
	 \Aitem \While k $<$ Nb  
	   \Aitem \mt \For n = 1 \To Nb
	 	\Aitem \mttt e0 $\setto$ pe(n)
	 
		 \Aitem \mtt \For i = r\_index + 1 \To Nh  
	 		\Aitem \mtttt r(i) $\setto $ r(i)+ e0 * h(i - r\_index)
	 
	 \Aitem \mtt \For i = 1 \To r\_index  
	 \Aitem \mtttt r(i) $\setto $ r(i)+ e0 * h(Nh - r\_index + i )
	 
	 \Aitem \mtt sp(n) $\setto $ r(r\_index)
	 
	 \Aitem \mtt r(r\_index) $\setto$ 0
	 
	 \Aitem \mtt\If r\_index $<$ Nh
	 \Aitem \mtt \Then r\_index $\setto$ r\_index + 1
	 \Aitem \mtt \Else  r\_index $\setto$ 0
	 \Aitem \Return ps
\end{algorithm}
\end{figure}
%%%%%%%%%%%%%%%%%%%%%%%%%%%%%%%%%%%

\pa Within this pseudo code, some details can be underlined.

\pa We refind the fact that the current output sp(n) (linr 13) is available quite fastly, as soon as both "for" loops realizing the superposition of remaining residue (stored in r) and new tache are performed (lines 9 and 10, lines 11 and 12).

\pa These both "for" loops use a shift related to the index r\_index, index of the circular buffer r, which seems to confers its rapidity to this code.

\pa The update of r\_index is done, after having set to zero the current pointed value r(r\_index) (line 14), in the lines 15 to 17 with a control of the r\_index value in order to go back to the beginning of r as soon as its ends has been reached.

\pa Some preliminary tests using the STL library, which permits to vectorize the code,  have been done and if the resulting code is rather shorter and simpler, the needed computing time was really quite more important, unacceptable in fact. So, we have not considered a vector-calculation optimisation using, in our case for the moment, CoreAudio optimization. Indeed, the available time for the project is rather short and we want to propose cross platform versions of the applications, notably the IDS analyzer and re-synthesizer which is the core of the current project. The convolution problem is just a small part of this project. 

\pa Clues for optimization of the code are then the use of Simple Instruction Multiple Data (SIMD) sets of instructions according to the processor and operating system used, but it will prevent form portability of the code. Porting the code on dedicated DSP should be considered with care because of the potential cost of such an operation.

\pa A first optimization rather simple to perform would be to introduce a test to bypass the update of the Buffer, that is to say both "for" loops (lines 9 to 10 and 11 to 12) either when the running input value, pe(n), is zero or when it would be interesting to enforce its value to zero. It may be interesting because we can find many '"silent" parts in pieces of music or speech but it will induce some choices to determine the minimum level to link to the zero value.

\pa Another clue is related to the vector-base code, assuming the use of vectors for at least impulse response, residue r and output signals.  Indeed, it could be interesting to investigate the use of Graphic Processing Unit (GPU) as they should be optimized to perform fast operations over matrices and we can organize the data for a given channel as a part of a matrix. The key factor would be the  calculation precision available for GPUs. But it could be a promis\-sing track, using low level code or optimized librairies as CoreImage on Mac Os X platform for instance.

\pa The source code of a read-to-compile cross platform will be available on a dedicated web site, with also versions of the IDS analyser and re-synthetizer before the end of the year. This codes will notably manage the choice, reading and writing of useful sound files: input, impulse response and output. 

\pa For more information about this subject, please mail to the authors after september 2008, as we may build the web site at the end of the project. 
%%%%%%%%%%%%%%%%%%%%%%%%%%%
\section{Taches-algorithm performance}

\subsection{Influence of filter impulse response size}

\begin{table*}[h!t!]
\begin{center}
\begin{tabular}{|l|l|l|c|c|}
\hline
Computer model & name & Available RAM & MacOs & Xcode \\
\hline
PowerBook 1.67 GHz PowerPC G4 & PBG4 & 2 Go DDR2 SDRAM & 10.4.11 & 2.5\\
PowerMac 2x 2.3 GHz PowerPC G5& PMG5 & 4 Go DDR SDRAM & 10.4.11 & 2.5\\
MacBook Pro 2.4 GHz Intel Core 2 Duo & MBP &  2 Go 667 MHz DDR2 SDRAM & 10.4.11 & 2.5\\
MacBook 2.2 GHz Intel Core 2 Duo & MB & 2 Go 667 MHz DDR2 SDRAM & 10.5.1 & 3.0\\
MacPro 2x2 2.8 GHz Quad-Core Intel Xeon & MP8 & 4 Go 800 MHz DDR2 FB-DIMM & 10.5.2 & 3.0\\
\hline
\end{tabular}
\caption{Information about the computers used for the experiments, notably the name used in the following.}
\label{tab:Ordis}
\end{center}
\end{table*}

\pa For each tested computer, we used three stimuli of different length (30~s, 67~s and 129~s) and study the behavior of the same Xcode  C++ project, compiled on each computer, permitting the implementation of the taches-algorithm. The code used is then the one which does not rely on vector-based operations, but, buffer-based ones.

\pa The studied parameter is the filter impulse response length in order to determine which is the limit filter length permitting, \textit{a priori}, a real-time convolution using the taches-algorithm instead of the fast convolution using FFT and IFFT. To consider a rough case, all the used impulse res\-ponses are designed using white noises of chosen durations, generated by Audacity sound editor. With such impulse responses, we tried to be in very time consuming convolutions for each experiments, maybe similar to some worst convolution cases.

\pa The tested computers are defined in Tab.~1, giving some information about the processor(s), their number, the RAM available, the version of the operating system used and the version of the Xcode IDE we used. Indeed, even if we have been working on cross platform code, for the moment the whole developement and the tests have been made only under Mac\-OS X, using the native IDE, Xcode. 

\pa We intend to keep on with performing the same set of tests for Intel based computers either using the BootCamp Application on Apple computers or "real" PC computers with notably Intel processors and Windows XP or Vista.  

\pa For each computer tested, the plots proposed in the following figures have three main curves which correspond to the three musical stimuli used:
\begin{itemize}
\item square markers used for the shortest input file tested (30~s);

\item down triangle markers for the intermediate long input file (67~s),

\item up triangle markers for the longest input file (129~s).
\end{itemize} 
 
\pa For adapted filter size ranges, that is to say not the whole range for slowest computers, three horizontal segments are plotted which respectively correspond to the duration of each stimuli. Then, by looking at the intersection between related calculation time and  stimulus horizontal curves, we can determine a limit length which may constitute a potential indicator of the longest filter size which would be used to perform real-time convolution on the studied computer. Indeed, with the test program we used, we performed impulse response and input signals reading, convolution calculation, output signal writing and calculation of several indicators in order to be able to plot the curves.
 
\pa But, in order to design a version potentially able to perform real-time convolution, with the taches-algorithm, we would not have to introduce the calculation of control indicators and data reading or writing. Management of the input, impulse response and output data would be realized by another tool(s). In order to verify if the filter limit lengths are valid, we have been working on convolution dedicated demonstrators to be presented during the Convention, and, notably pure data externals. 
 
\pa Fig.~7, 8 and 9 present the calculation times we get using the three Intel based computers: MB, MBP and MP8. 

\begin{figure}[h!t!]
\begin{center}
\includegraphics[width=12cm]{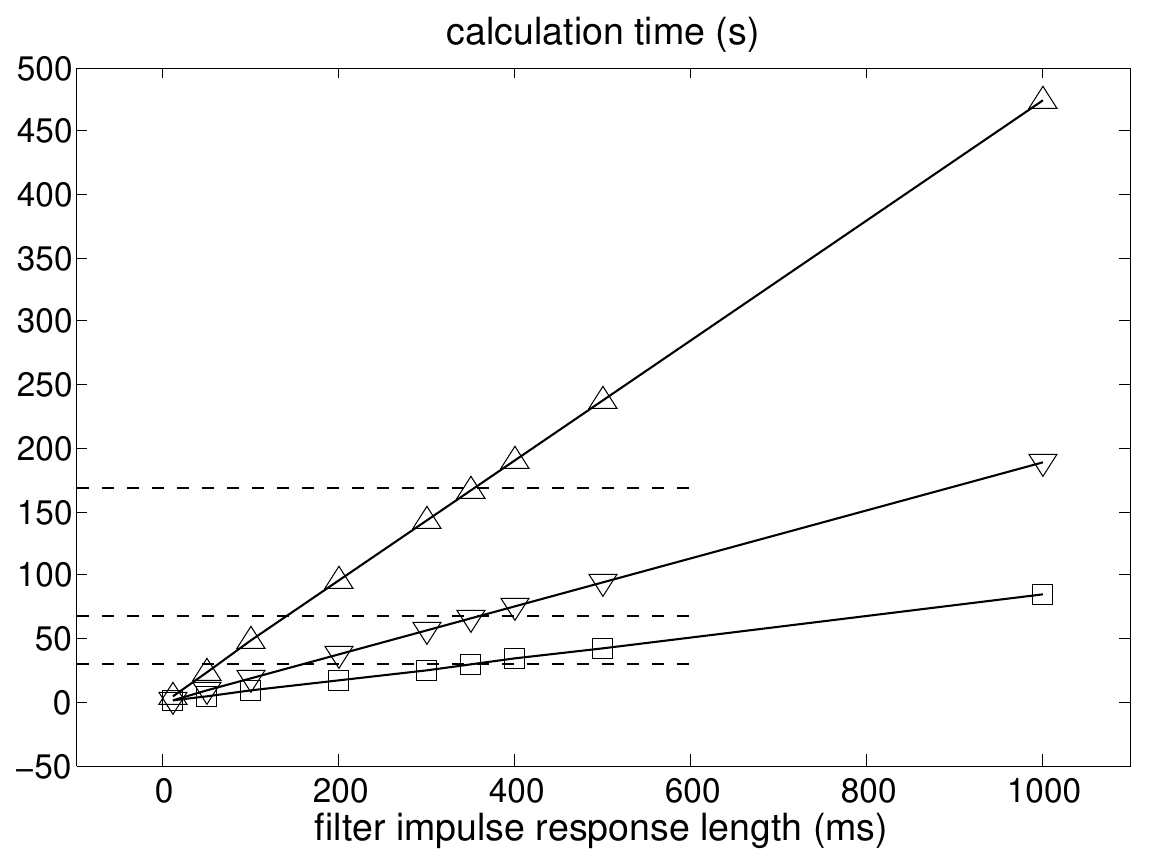}
\end{center}
\nobreak
\caption[7]{Influence of the filter impulse response length for the MB computer. Three horizontal segments respectively correspond to the duration of each studied stimulus, so that we can determine the \textit{a priori} limit size for real-time convolution using the taches-algorithm for the tested computer: this limit impulse response size may correspond to the intersection between the horizontal segment and the related calculation time curve.}
\end{figure}

\begin{figure}[h!]
\begin{center}
\includegraphics[width=12cm]{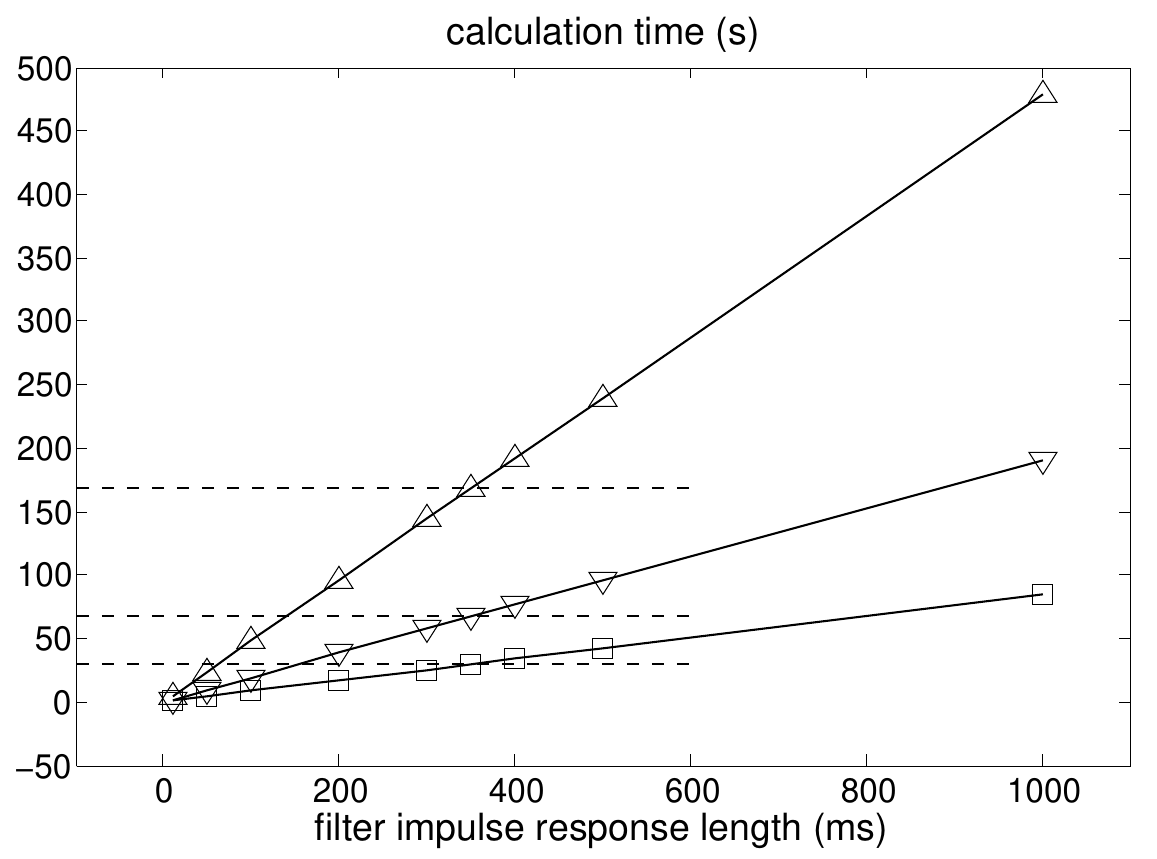}
\end{center}
\nobreak
\caption[8]{Influence of the filter impulse response length for the MBP computer. Three horizontal segments respectively correspond to the duration of each studied stimulus, so that we can determine the \textit{a priori} limit size for real-time convolution using the taches-algorithm for the tested computer: this limit impulse response size may correspond to the intersection between the horizontal segment and the related calculation time curve.}
\end{figure}

\begin{figure}[h!]
\begin{center}
\includegraphics[width=12cm]{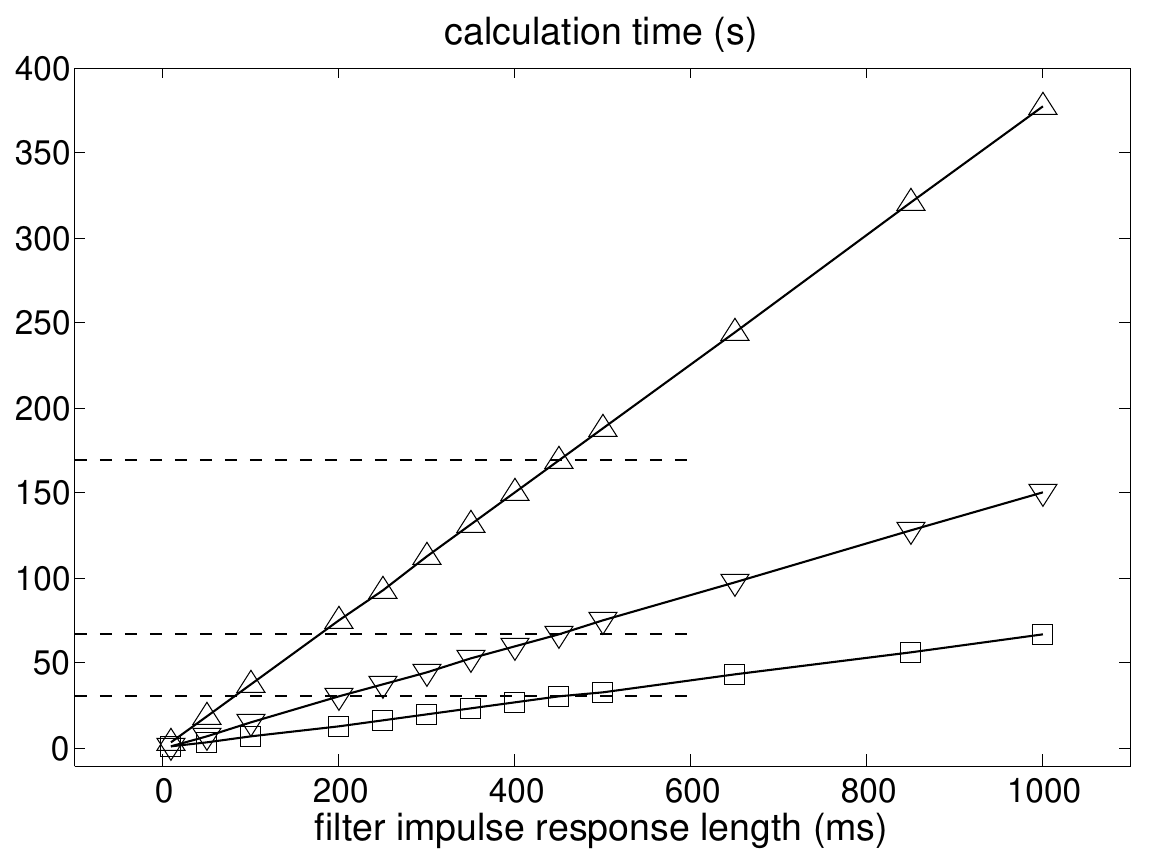}
\end{center}
\nobreak
\caption[9]{Influence of the filter impulse response length for the MP8 computer. Three horizontal segments respectively correspond to the duration of each studied stimulus, so that we can determine the \textit{a priori} limit size for real-time convolution using the taches-algorithm for the tested computer: this limit impulse response size may correspond to the intersection between the horizontal segment and the related calculation time curve.}
\end{figure}

\pa The results for MB and MBP are rather similar even if the MBP may be a faster processor but, during the experiments, the MBP computer was  using an older version of MacOS X (10.4.11 instead of 10.5.1 for the MB), and of Xcode (2.5 instead of 3.0 for the MB). We think that using the last version of MacOs X and Xcode, the MBP computer could be a little bit faster than the MB one. 

\pa The limit filter size seems to be close to 350~ms for the MB computer   and for the MBP computer (using MacOs 10.4.11 and Xcode 2.5), which means that the limit filter would increase a little for the MBP computer due to the optimization of the Operating System and of the Xcode application. For the MP8, this limit filter size seems close to 450~ms and may be higher with a code taking into account the presence of 8 processor cores if it is efficient to dedicate parts of the processing to several cores. But, it would be also interesting to think about using the 8 cores to realize convolution of multichannel stimuli.  

\pa We can underline the fact that the behavior of the algorithm is linear, for all the stimuli, up to 1~s filter duration.  It would be interesting to perform the tests for longer impulse responses to get an idea of the behavior of each computer when the amount of calculation is really increased, which should be done in future after some further code optimizations.  

\pa Fig.~10 and 11 present the global evolutions of calculation times for both PowerPC based computers: PBG4 and PMG5.
 
\begin{figure}[h!t]
\begin{center}
\includegraphics[width=12cm]{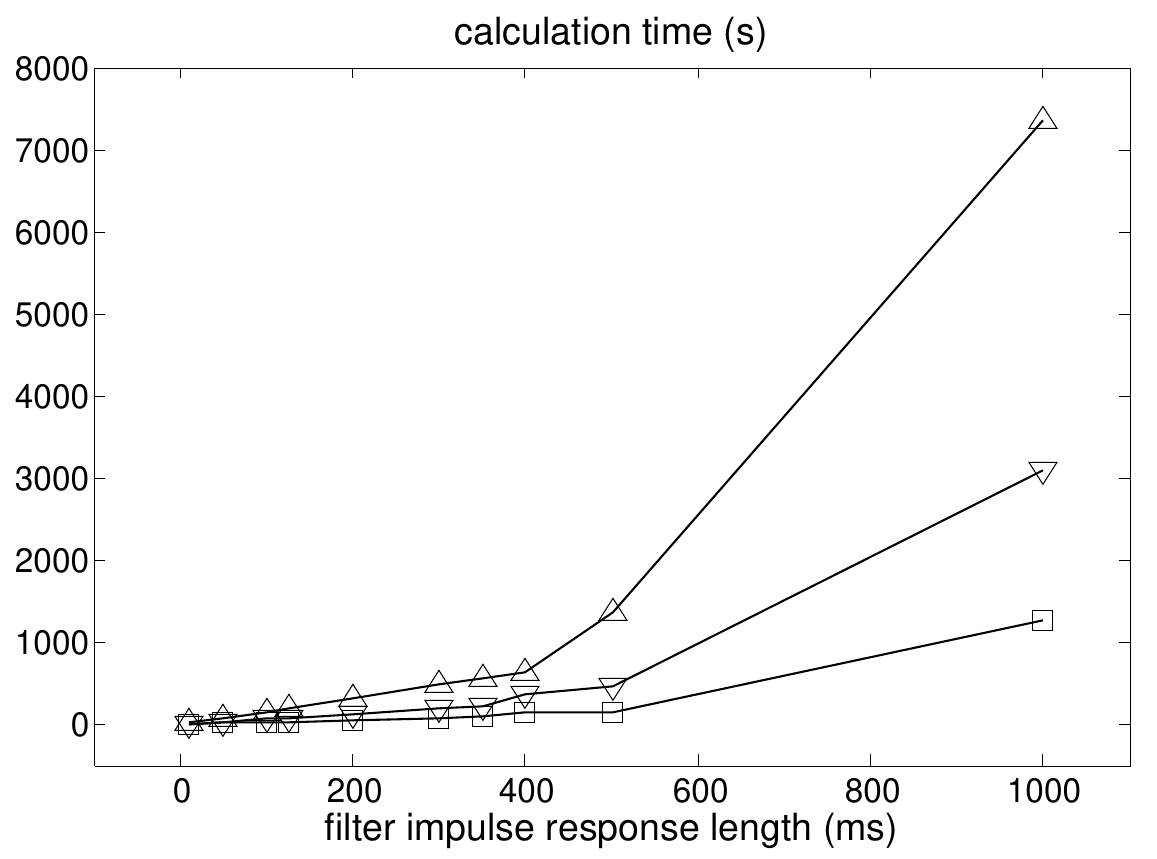}
\end{center}
\nobreak
\caption[10]{Influence of the filter impulse response length for the PBG4 computer. Due to the huge variation of the calculation time for the range of tested impulse response lengths, we only plot the calculation time related to each stimulus for the whole filter lengths range.}
\end{figure}

\begin{figure}[h!t]
\begin{center}
\includegraphics[width=12cm]{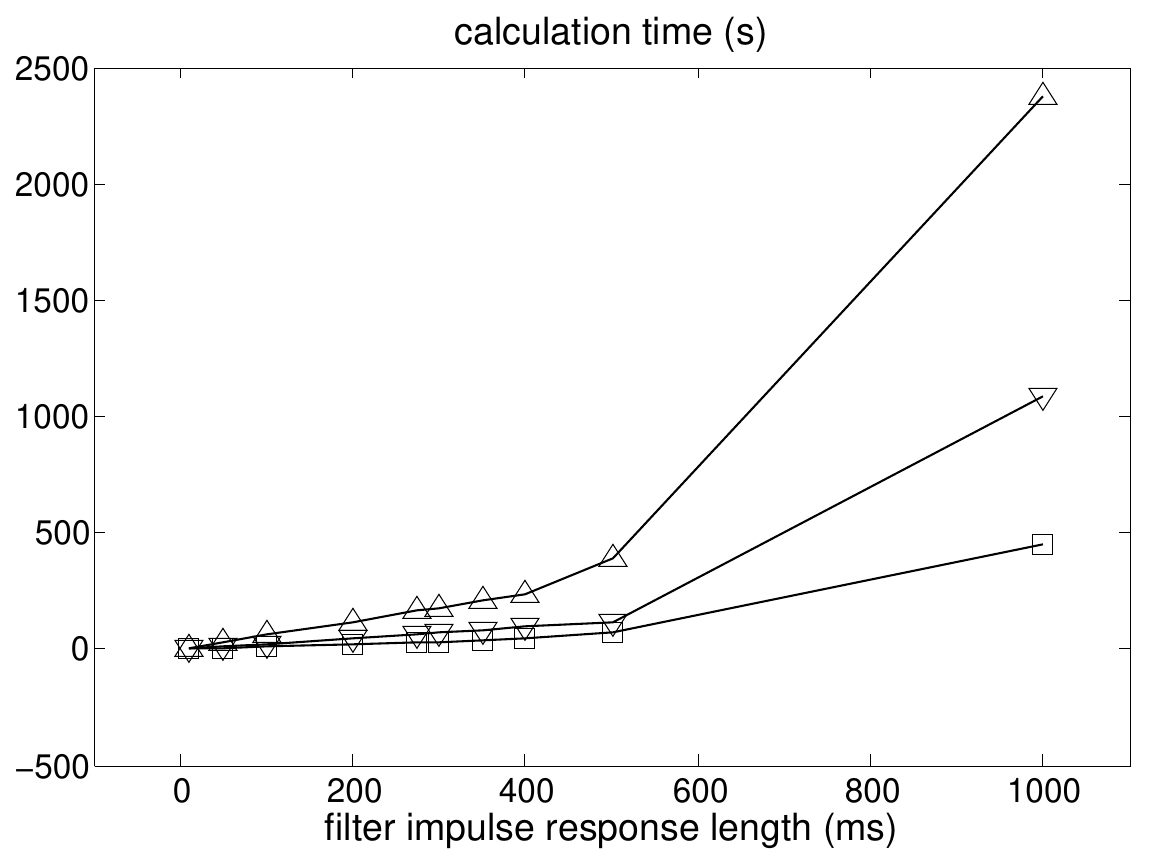}
\end{center}
\nobreak
\caption[11]{Influence of the filter impulse response length for the PMG5 computer. Due to the huge variation of the calculation time for the range of tested impulse response lengths, we only plot the calculation time related to each stimulus for the whole filter lengths range.}
\end{figure}

\pa For both computers the figures for the calculation times over the whole filter lengths range  show that the computing power is slower than for Intel based computers. The gaps are rather large so we do not include the plots of the stimuli durations. We can note that the behavior is not linear over the whole length range and seems to increase with the size of the considered stimulus. As these compu\-ters are at least two year older than the three Intel based others it seems normal to observe an increase of the computing power for recent computers, but, this increase can also be related to the use of a different family of processors: Intel ones instead of PowerPC G4 or G5 one. This is a promising result which means that, in future, the limit filter size would be higher and higher, with a potential linear behavior over an extended filter length range allo\-wing fast convolution even for off-line processing using the FFT-free taches-algorithm.

\pa Fig~12 and 13 permit to determine the limit filter size for both PowerPC based computers.   

\begin{figure}[h!t]
\begin{center}
\includegraphics[width=12cm]{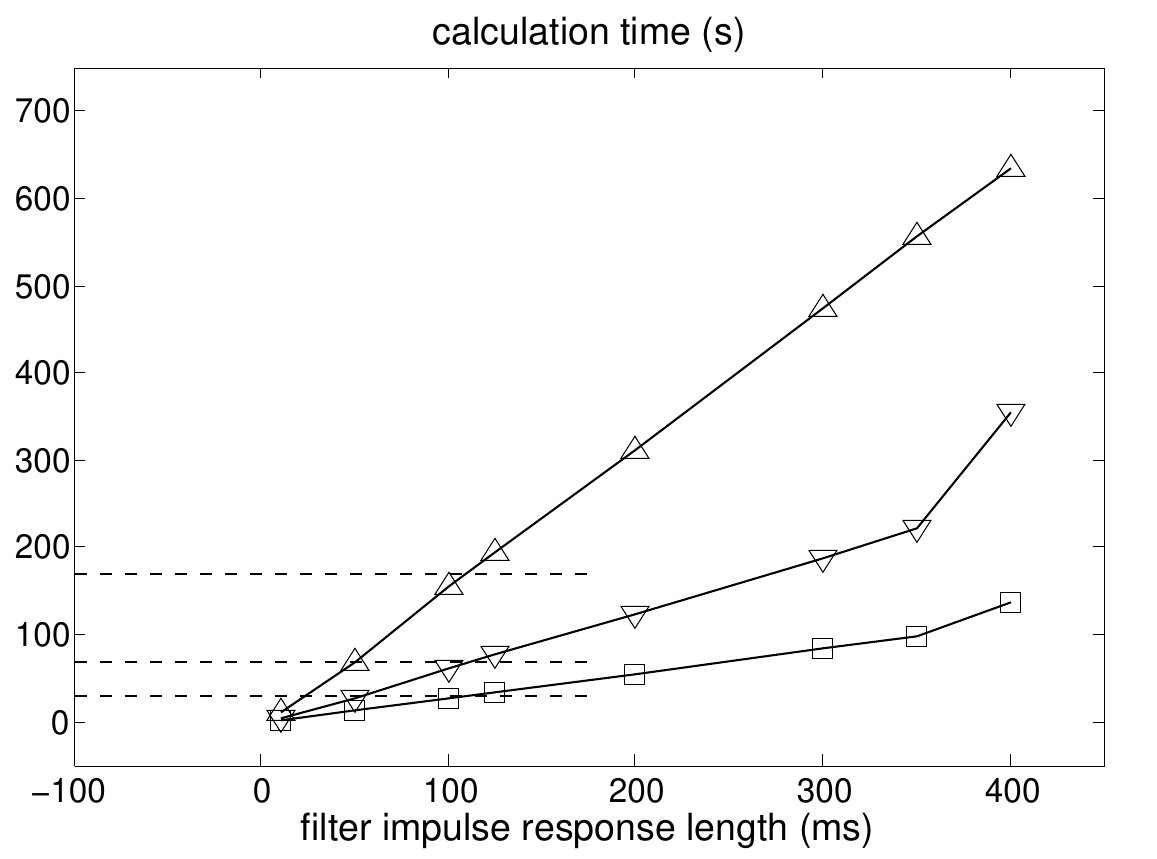}
\end{center}
\nobreak
\caption[12]{Influence of the filter impulse response length for the PBG4 computer for the useful range of tested impulse response lengths. Three horizontal segments respectively correspond to the duration of each studied stimulus, so that we can determine the \textit{a priori} limit size for real-time convolution using the taches-algorithm for the tested computer: this limit impulse response size may correspond to the intersection between the horizontal segment and the related calculation time curve.}
\end{figure}

\begin{figure}[h!t]
\begin{center}
\includegraphics[width=12cm]{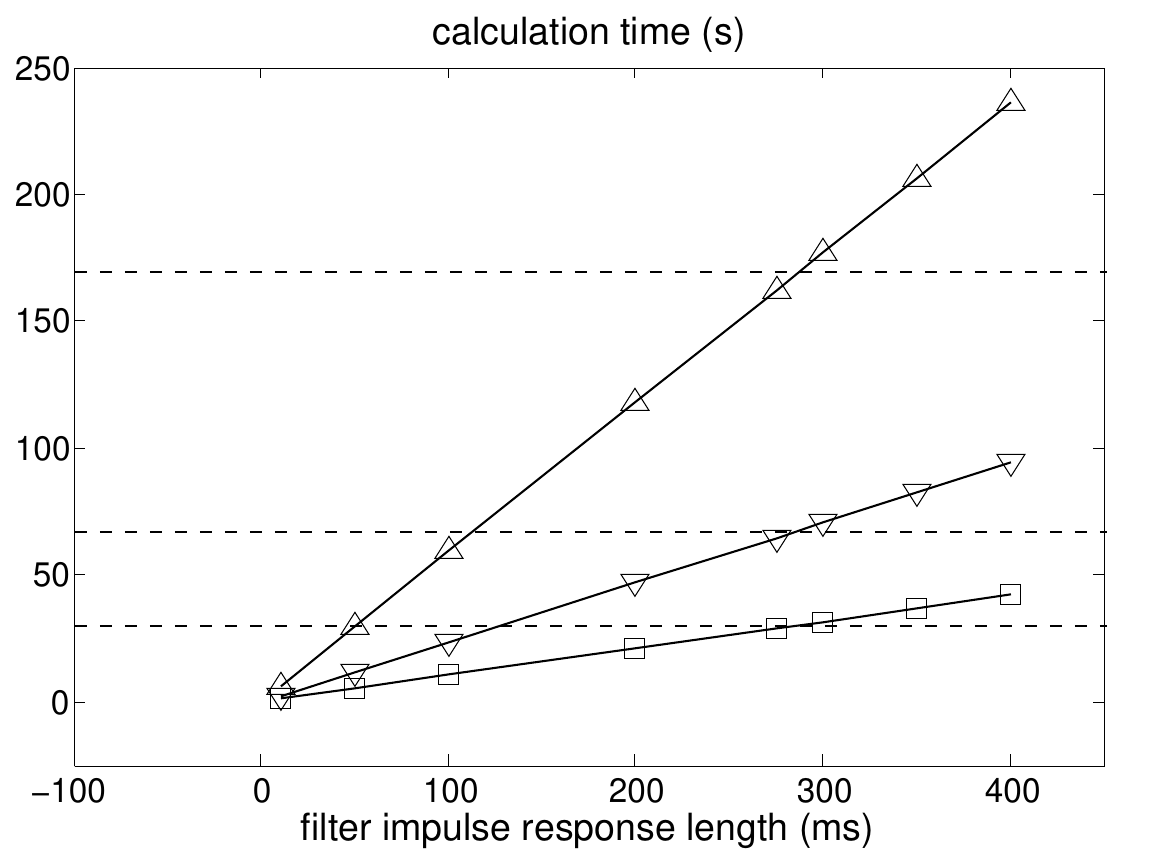}
\end{center}
\nobreak
\caption[13]{Influence of the filter impulse response length for the PMG5 computer for the useful range of tested impulse response lengths. Three horizontal segments respectively correspond to the duration of each studied stimulus, so that we can determine the \textit{a priori} limit size for real-time convolution using the taches-algorithm for the tested computer: this limit impulse response size may correspond to the intersection between the horizontal segment and the related calculation time curve.}
\end{figure}

\pa For the slowest, the PBG4 computer this limit filter size seems close to 125~ms while, for the PMG5, it seems close to 275~ms, values 75~ms to 175~ms shorter than the ones found for the three Intel based computers. And, the behavior is linear for the three stimuli up to 350~ms for the PBG4 computer and 400~ms for the PMG5 one.

\subsection{Influence of input buffer size}
\pa To test the influence of the input buffer size we have only really considered the case of the MB computer, with MacOs 10.5.1 and Xcode 3.0 as it may give an idea for a basic computer in mobile music applications needing convolution.  

\pa We do not give the results for the MP8 computer, even if one may think to find such a machine in studio, because the code used for the test is not optimized to take into account the 8 cores and should be optimized in such a way to be representative. But we can say that we found the same trends for the MP8 computer compared to the MB.  

\pa We did not realized the whole tests for the MBP computer as we thought it would be more interesting to perform such a test when the upgrade to MacOS 10.5.1 and Xcode 3.0 can be done. 

\pa The input buffer size corresponds to the Nb parameter defined in the  buffer-based code and represents the maximal number of samples constituting a block for the "while" loop. The number of samples would be often lower for the last block as the input duration is seldom a multiple of the input buffer size. We tested powers of two Nb values to have an element of comparison for current block fast convolution using FFT and IFFT. For the tests of the influence of the filter length, Nb was set to 8192 samples which is already a consequent block size but which represents, as explained in the paragraph, a good trade-off. 

\pa In order to be able to notice some differences we considered the case of the longest stimuli (169~s) and of 350 (limit filter size for the MB computer) and 500~ms. The results are given in Tab. 2.

\begin{table*}%[h!]
\begin{center}
\begin{tabular}{|l|r|r|r|r|r|r|r|r|r|r|}
\hline
Nb & 256 & 512 & 1024 & 2048 & 4096 & 8192 & 16384 & 32768 & 44100 & 65536 \\
(samples) &  & & &  &  &  &  &  &  &  \\
\hline
h: 350 ms &  166.69 & 166.50 & 166.43 & 166.41 & 166.31 &166.33 & 166.29 & 166.29 & 166.26 &  166.36 \\
h: 500 ms & 237.83 & 237.51 & 237.34 & 237.42 & 237.29 & 237.28 & 237.38 & 237.19 & 237.14 & 237.18 \\
\hline
\end{tabular}
\caption{Influence of the input buffer size, Nb parameter in the table, for the longest stimulus (169~s) and 350 and 500~ms lengths for the impulse response.}
\label{tab:BUF_LEN}
\end{center}
\end{table*}

\pa Considering the results, one can easily note that differences between  
calculation times for either the limit filter size of 350~ms or the filter size of 500~ms have respectively a magnitude of 33~ms and 63~ms, which correspond to tiny differences while we either multiply the reading operations associated with small arrays in memory for short Nb va\-lues or have in memory quite huge arrays for a little number of reading operations for long Nb values.

\pa To perform tests over various computers, the choice of a Nb value of 8192 samples seems a rather good compromise.

\pa But, the tiny influence of the Nb para\-meter is rather surprising as we have tested values even greater than one second!
 
\subsection{Question of the requantization}
\pa We present in this paragraph the estimated requantizations, using the taches-algorithm, considering two different assumptions: 
\begin{itemize}
\item the theoretical case of an infinite dynamic for the accumulator and then only one requantization operation;
\item the case of a requantization operation introduced after each sum of two scaled values, the so-called multiply and accumulate operation (MAC).
\end{itemize}

\pa We also derive our results considering data in fixed-point format using $N$-bit word-length. Extension to numbers in floating-point format should need some further work...

\pa For both estimations, we consider the following mathematical formula to derive the output signal:
 $$s[n]=\sum_{m=0}^{N_h-1}e[m].h[n-m]$$
where $N_h$ corresponds to the length of the impulse response signal or tache. 
 
 \subsection{Ideal requantization}
\pa We consider the case where we have an infinite dynamic for the accumulator and we perform the requantization operation at the end of the whole calculation for each sample of the output signal. 
 
\pa  In such a theoretical case, we can note that to calculate the running output sample $s[n]$, we need to sum $N_h$ products $e[m].h[n-m]$. 
 
\pa If we assume that input and impulse response samples are coded using $N$ bits, each product needs $2N$ bits to be correctly represented. And, as we have $N_h$ products in the sum, each output sample would need $2N+N_h-1$ bits for a correct representation.

\pa So, for the theoretical case, we would have a single requantization which removes $N+N_h-1$ bits. 
 
\subsection{"MAC" requantization}
\pa In this case, we consider that we make a requantization operation after each MAC operation.

\pa Considering the mathematical formula, we can note that we must sum 
$N_h$ scaled values of either $e[m]$ or $h[n-m]$ according the adopted point of view. This means that we must perform, in fact, $N_h-1$ MAC operations with each MAC operation related to a requantization suppressing $N$ bits.

\pa Then in the "Mac" requantization scheme, we would have to perform $N_h-1$ requantization operations removing $N$ bits.

\pa For both requantization assumptions the number of necessary requantizations seems quite lower than the one needed even for the calculation of a DFT while the fast convolution needs two DFTs, a point-by-point multiplication in frequency-domain and a final IDFT. And, even one DFT involves more operations than the number of operations needed for the whole calculation of one sample of the output using the taches-algorithm.

\pa Thus, we think that the taches-algorithm may introduce quite less requantization error compared to the DFT fast convolution and, probably even compared to the fast convolution using FFTs and IFFTs. 

\pa We then strongly consider the taches-algorithm as a quite precise convolution alternative, for instance for measurements.

\section{Potential uses of the taches-algorithm}
\pa Considering the question of requantization, we suggest to consider the taches-algorithm for off-line measurements methods. And, we have been using it as convolution algorithm within the cross platform C++ porting of the existing Scilab prototype of the IDS analyzer developed by Millot \cite{Millot:04, Millot:07} which would permit to propose physical phenomenon characterization, audio installations cha\-racterization, audio scenes analysis and cartography, for instance, using a perceptive frequency mapping and broadband excitation stimuli: white or pink noise but also pieces of music, speech, ambient noise, ... Extension to analysis of medical signals is also on duty, using the IDS analyzer and the taches-algorithm.

\pa Compared to fast convolution, the taches-algorithm is also a credible alternative as the calculation of the running output sample only depends on the optimization of the taches superposition operation. With actual computers, we may be able to perform real-time convolution with very low latency, using the taches-algorithm, for impulse response signals as long as at least 350~ms to 450~ms without any SIMD optimization, for the moment. Thus, introducing such optimizations we would be able to perform real-time convolution for longer impulse response signals and, using the multi cores, probably for multichannel signals.

\pa Another advantage of the taches-algorithm is the fact that we can think about changing the nature or behavior of the impulse response even sample-by-sample while we would need at least the calculation of an input block for the fast convolution. It would then be possible to modify in real-time the beha\-vior of given room acoustics frequency range by frequency range, changing decay and/or reverberation noise within frequency bands, or changing progressively either the early or late reflections temporal location and influence which may give alternative ways to perform hybrid reverberators and cope with analysis problems concerning the impulse response in frequency domain \cite{Stewart:07}.  

\pa This may be useful for situations where either the source(s) or the receiver move(s) inside given complex acoustics, that is to say for instance within video games,  videos or movies. It could also be quite interesting in real-time interaction, during a concert, between musicians and a person behind a mixing desk with taches-algorithm processors: for instance to use the recording of one instrument to filter another one and spatialize it. Or, within an interactive installation, it could be used to take into account the moves of people inside the installation to modify the behavior of the installation from a visual, video or sound point(s) of view.

\pa It could also be used in real-time recording simulators \cite{Braasch:05, Millot:06} to introduce time-varying room acoustics in such anechoic and free-field simulators. 

\pa We also think about using it in order to perform off-line and, above all, real-time cartography \cite{Millot:05} within complex audio scenes with the trajectories of the emerging or main sources being plotted on a screen to follow their moves.  
 
\pa And, obviously, it may be able to realize real-time convolvers, correlators or perceptive sound meters using IDS perceptive decomposition as plugins or PDA components.

\section{Conclusion}
\pa With this paper, we intended to present another way to think and compute convolution in time-domain.

\pa We have proposed some clues and pseudo codes to realize such time-domain convolution and demons\-trated the power of the proposed algorithm.

\pa The study of the influence of the filter length and of the requantization related to this algorithm seems quite promising either to realize high quality off-line measurements tools and audio processors or even real-time processors using "medium" filter lengths at least.

\pa Various applications for the taches-algorithm can be considered either in off-line or real-time.

\pa If we have some ideas to optimize the current version of the taches-algorithm we would not be able to perform them alone and quickly. So, we would provide the C++ code source as soon as possible with also ready to compile versions of the taches-algorithm and of the IDS analyzer. Interested people would then be able to test these tools by themselves in order to determine whether they are or not as interesting as we think and/or to help us to propose better and optimized versions of these tools. We also have been working in order to propose, at least, pure data externals implementing the taches-algorithm in real-time objects for easy and flexible time-domain convolution.

\section{Acknowledgements}
\pa The authors would like to thank the following people for the support of past and current research works:
\begin{itemize}
\item Samuel Tracol, former student at the FSMS and currently working at Flux:: (www.fluxhome.com), who worked on the first porting of the taches-algorithm in C (2003);
\item Henry Bernard and Fabien Lepercque who have been currently working on the portage on the IDS analyser and re-synthesizer and helped us to realize the tests of the taches-algorithm power and code the Xcode version of the test program;
\item Cap Digital (www.capdigital.com)  which has given funds in order to realize the cross platform portage of the IDS analyser and re-synthesizer and propose an alternative cha\-racterization method for audio installations devoted to videos and movies production as parts of the HD-3D IIO project.
\end{itemize}

%%%%%%%%%%%%%%%
\end{document}